\def\ltsim{\mathrel{<\kern-1.0em\lower0.9ex\hbox{$\sim$}}}          %personal def
\def\gtsim{\mathrel{>\kern-1.0em\lower0.9ex\hbox{$\sim$}}}          %personal def
\documentclass{aa}
\begin{document}
\thesaurus{03(11.01.2; 11.09.4; 13.09.1; 13.09.3; 13.18.1)}
\title{ ISO observations of a sample of Compact Steep Spectrum and
GHz Peaked Spectrum Radio Galaxies}
\author{C. Fanti\inst{1,2} \and F. Pozzi\inst{3,2}\and R. Fanti\inst{1,2}  
\and S.A. Baum\inst{4} \and C.P. O'Dea\inst{4} 
\and M. Bremer\inst{5}
\and D. Dallacasa\inst{3,2}
\and H. Falcke\inst{6}
\and T. de Graauw\inst{7}
\and A. Marecki\inst{8}
\and G. Miley\inst{9}
\and H. Rottgering\inst{9}
\and R.T. Schilizzi\inst{10}
\and I. Snellen\inst{11}
\and R.E Spencer\inst{12}
\and C. Stanghellini\inst{13}
}                     
\offprints{F. Pozzi}                              
\mail{cfanti@astbo1.bo.cnr.it}
\institute{Dipartimento di Fisica, Universit\`a di Bologna 
Via Irnerio 46, I--40126 Bologna,Italy 
\and  Istituto di Radioastronomia del CNR, Via Gobetti 101, I--40129 Bologna, 
Italy
\and    Dipartimento di Astronomia, Universit\`a di Bologna, via Ranzani 1 
I-40127 Bologna, Italy 
\and STScI, 3700 San Martin Dr., Baltimore, MD 21218, USA 
 \and Department of Physics, Bristol University, H H Wills Physics
Laboratory,  Tyndall Avenue, Bristol, BS8 1TL, U.K.
 \and Max-Planck-Institut f. Radioastronomie, Auf dem Huegel 69, D-53121
Bonn, Germany 
\and  Laboratorium voor Ruimteonderzoek, SRON, Postbus 800,  NL--9700 AV
Groningen,The Netherlands 
 \and Toru\'n Centre for Astronomy, N. Copernicus University, ul. Gagarina 11,
PL-87-100 Torun, Poland
\and  Sterrewacht, Oort Gebouw, P.O. Box 9513,  NL--2300 RA Leiden, The
Netherlands 
\and Joint Institute for VLBI in Europe, Postbus 2, NL--7990 AA, Dwingeloo The
Netherlands   
\and  Institute of Astronomy, Madingley Road, Cambridge CB3 0HA, UK 
\and Jodrell Bank Observatory, University of Manchester, Macclesfield, 
Cheshire SK 11 9L, UK 
\and Istituto di Radioastronomia del CNR, C.P. 141, I-96017 Noto (SR), Italy
}                                        
  \date{Received December 21, 1999 / Accepted April 10, 2000} 
% must be given  at the last moment
%command, it will format your titlepage.
%
\maketitle 
\markboth{FIR properties of CSS/GPS radio galaxies}{Fanti et al.}
\begin{abstract}
We present results from observations obtained with ISOPHOT, on board the ISO
satellite\footnote {Based on observations with ISO, an ESA project with
instruments funded by ESA Member States (especially the PI countries: France,
Germany, the Netherlands and the United Kingdom) with the participation of
ISAS and NASA.}, of a representative sample of seventeen CSS/GPS radio galaxies
and of a
control sample of sixteen extended radio galaxies spanning similar ranges
in redshift ($0.2 \leq z \leq 0.8$) and radio luminosity ($P_{2.7 {\rm GHz}} \geq
10^{26}$ W/Hz).
The observations have been performed at $\lambda$ = 60, 90, 174 and 200 $\mu$m. 
The original purpose of these observations was to check whether CSS/GPS sources 
are
associated with very gas rich galaxies, as required by the scenario in which 
the growth of the radio source is inhibited by the dense medium of the
host galaxy.

Unfortunately the resulting performance of ISOPHOT was  worse than expected. 
As a consequence, the detection limit at 60$\mu$m is similar to that obtained 
previously with IRAS but better than that at 90$\mu$m.

Seven of the CSS/GPS sources have detections $\geq 3\sigma$ 
at one or more wavelengths, one of which is detected  at $\geq 5\sigma$. 
For the comparison sample five objects have detections $\geq 3\sigma$ one of
which is at $\geq 5\sigma$.

By co--adding the data we have obtained average flux densities 
at the four wavelengths.

We found no evidence that the FIR luminosities of the CSS/GPS sources are
significantly different from those of the extended objects  and therefore 
there is not any support for CSS/GPS sources being objects ``frustrated" by 
an abnormally dense ambient medium.

The two samples were then combined, providing FIR information on a new
sample of radio galaxies at intermediate redshifts. We compare this
information with what previously known from IRAS and discuss the average
properties of radio galaxies in the redshift range 0.2 -- 0.8.
The FIR emission cannot be accounted for by extrapolation of the synchrotron
radio spectrum and we attribute it to thermal dust emission. The average FIR
luminosity is
$\geq 6 \times 10^{11} L_{\odot}$. Over the observed
frequency range the infrared spectrum can be described by a power law with
spectral index $\alpha \simeq 1.0 \pm 0.2$. Assuming the emission to be due to
dust, a range of temperatures is required, from $\geq 80$ K to $ \approx 25$
K.
The dust masses required  to explain the FIR emission range from $5 \times 10^5 ~ M_{\odot}$ for the 
hotter component up to $ 2 \times 10^8~M_{\odot}$ for the colder one. 

We present also observations on four nearby ($z \leq 0.1$) GPS radio galaxies, two of which 
are detected at all four wavelengths.

\end{abstract}
\section{Introduction}
Compact Steep Spectrum (CSS) and GHz Peaked Spectrum (GPS) radio sources are
powerful extragalactic radio sources with radio emission confined well within
their host galaxy/quasar ($\ltsim 15 $kpc)\footnote{Throughout this paper we 
use  H$_0$=100 (Km/sec)/Mpc and q$_0$=0.5.}. They are as powerful as the 
classical FRII radio sources but are of much smaller physical size and yet have  
normal/steep radio spectra at GHz frequencies. 
A discussion of the properties of CSS can be found in Fanti et al.
(\cite{fanti}; \cite{fanti2}). A comprehensive and updated review on CSS and GPS is presented 
by O'Dea (\cite{odea3}).

\begin{table*}
\begin{center}
\caption{CSS/GPS sample }
\begin{tabular}{|lllrl|rrrr|} \hline
      &       &                &   &       & \multicolumn{4}{|c|}{nominal
observing time}\\
Source& z\quad& LogP$_{2.7{\rm GHz}}$& LS& Sample& 60$\mu$m& 90$\mu$m& 174$\mu$m&
200$\mu$m\\
& & (W/Hz)\qquad& kpc& & sec\quad&sec\quad &sec\quad &sec\quad \\ \hline

\object{0108+38} &  0.67 &   26.79 & 0.023  & PW & 256 & 64 & 128 & 128      \\
\object{3C49}    &  0.62 &   26.89 & 3.6    &3CR & 192 & 64 &  96 & 128     \\
\object{3C67}    &  0.31 &   26.30 & 6.8    &3CR & 256 & 128&   --- &  --- \\
\object{0404+76} &  0.60 &   27.23 & 0.53   & PW & 192 & 96 &  96 & 192 \\
\object{1244+49} &  0.21 &   25.63 & 5.7    &San & 256 & 64 & 128 & 128\\
\object{3C268.3} &  0.37 &   26.54 & 3.9    &3CR & 256 & 128&  32 &  32\\
\object{1323+32} &  0.37 &   26.75 & 0.18   & PW & 192 & 64 &  96 &  96\\
\object{1358+62} &  0.43 &   26.81 & 0.16   & PW & 64  & 32 &  32 &  32\\
\object{3C303.1} &  0.27 &   25.60 & 5.0    &3CR & 192 & 96 & 32  &  32\\
\object{1607+26} &  0.47 &   26.92 & 0.18   & PW &  64 & 32 &  32 &  32\\
\object{1622+66} &  0.20 &   25.66$^a$& 0.0013       & Sn &  64 & 64 &  64 & 128\\
\object{3C343.1} &  0.75 &   27.24 & 1.3    &3CR &   --- &  --- &  32 &  32\\
\object{1819+39} &  0.798$^b$ &   27.23 & 3.4    &PW  & 192 & 96 &  96 &  96\\
\object{1819+67} &  0.22 &   25.06 & 0.02       &Sn  & 192 & 64 &  32 &  64\\
\object{1829+29} &  0.60 &   26.96 & 9.3    &PW  & 64  & 32 &  64 & 128\\
\object{2342+82} &  0.74 &   27.25 & 0.66   &PW  & 192 & 64 &  64 & 128\\
\object{2352+49} &  0.24 &   26.18 & 0.15   &PW  &  64 & 32 &  64 & 128\\ \hline\hline
median$^d$ &   0.43 &   26.79 & 1.0    &       & & & &\\ \hline\hline
\object{3C318}   &  1.574$^c$ & 27.71 & 8.2    &3CR &  64 & 32 & 128 & 192\\ 
        &           &       &        &    &     &    &     &    \\
\object{1345+125}&  0.12 &   25.79 & 0.11   & St & 64  & 32 &  32 &  64\\
\object{1718-649}&  0.014&   24.20$^a$  & 0.0014       &Tin & 256 & 64 & 128 & 128\\
\object{1934-63} &  0.18 &   26.65 & 0.14       &Tzi &  ---  &  --- &   --- &  64\\
\object{1946+70} &  0.10 &   25.00 & 0.04   &Sn  & 64  & 64 &  64 & 128\\ \hline
\end{tabular}
\label{sample}
\end{center}

NOTES:

$a)$ the spectrum of this GPS peaks at $\approx$3 GHz, and its
$P_{2.7{\rm GHz}}$ has been extrapolated from the spectrum at high frequencies 

$b)$ revised redshift by Vermeulen et al. (\cite{vermeulen})

$c)$ revised redshift by Willott et al. (\cite{willott})

$d)$ the median values are for the above representative sample ($0.2 \leq z \leq 0.8)$

Sample -- PW: Peacock \& Wall (\cite{peacock});
 3CR: Laing et al. (\cite{laing});
San: Sanghera (\cite{sanghera}); Sn: Snellen et al. (\cite{snellen});
St: Stanghellini et al. (\cite{stanghellini}; \cite{stanghellini2});
Tin: Tingay et al. (\cite{tingay}); Tzi: Tzioumis et al. (\cite{tzioumis})
\end{table*}

Some 15 to 25 \% of sources in a flux density limited sample, depending on the 
selection frequency, belong to this class. 
Many of them have radio spectra which show a flattening or a more marked 
turnover at frequencies between 50 MHz and a few GHz; those sources with a 
turnover at about 1 GHz are the GPS sources. The turnover is usually attributed 
to synchrotron self--absorption, but also free--free absorption (Bicknell et 
al. \cite{bicknell}) and induced Compton scattering (Kuncic et al.
\cite{kuncic}) have been
considered.

\medskip
Given that CSS and GPS sources are {\it physically small}, i.e. of
sub--galactic dimensions, it has been suggested that they are either 
{\it(1)} young objects ({\it youth} scenario, Phillips \& Mutel
\cite{phillips}), which have 
yet to develop extended radio lobes or {\it(2)} sources where 
the radio emitting plasma is trapped by an unusually dense interstellar medium 
({\it frustration} scenario, van Breugel et al. \cite{vanbreugel}).

In the latter hypothesis, CSS/GPS radio sources would be as old as those 
of larger size but their jets would spend their whole lifetime trying to 
escape, without success, out of the interstellar medium. This scenario 
requires a rather dense interstellar medium (average density
 $n_{\rm ISM} \gg 1$ cm$^{-3}$ and total mass within 1 kpc
 $M_{\rm ISM} > 10^8 M_{\odot}$; see De Young \cite{deyoung}; Fanti et 
 al. \cite{fanti2})
which is in the path of the radio source. Even if the host galaxy
contains a lot of gas, if this is distributed in a disk perpendicular to the 
radio source axis, it will have little effect on the radio source propagation,
therefore any ``frustrating'' gas has to be distributed over a large
volume.
Attempts to probe the different phases of this interstellar gas (e.g. optical 
observations of the Narrow Lines, polarization studies, X--ray emission)
have given, in our opinion, little support to the frustration scenario.

Recently, proper motion measurements of the hot spots in a few GPS
sources by Owsianik et al. (\cite{owsianik}) and Owsianik \& Conway
(\cite{owsianik2}) 
have shown separation velocities
$\geq 0.1~ c$. This finding  strongly suggests that these sources are  
very young. Also a recent study of the integrated radio spectra of a number 
of CSS/GPS (Murgia et al. \cite{murgia}) provides additional
evidence for short lifetimes.

It is nevertheless important to support these recent results with information 
on the ambient medium around these sources.

The frustration scenario requires that an unusually dense ISM be present 
in CSS/GPS sources. This may contain a substantial cold phase 
and a large amount of dust, and, in order to stop the advancing jet, this 
medium has to have a large covering factor. The dust will 
therefore absorb and re-process a fraction of the optical and UV radiation 
from the AGN (and young stars, if present) higher than in the larger size, 
``non--frustrated", radio sources, giving to the CSS/GPS sources an extra IR 
emission component in addition to the one from the disk/torus. 

Since this dense medium  has to extend over several kpc from the nucleus 
(in order to stop also radio sources 10--15 kpc in size), a large fraction of
the dust will be moderately cold so that the  emission is expected 
predominantly  in the medium--far IR (MFIR) spectral region. 

By searching for emission from cold dust, we have therefore tried to obtain
some constraints on this scenario. Since the emission peak is expected at 
relatively long wavelengths, ISO, with its capability of carrying out photometry at 
$\lambda \geq 60 \mu$m,  (Kessler et al. \cite{kessler}) was well suitable for such 
an investigation.

\medskip
This search is better suited to radio galaxies, rather than quasars, 
as in the former the contamination from the IR emission produced in the 
innermost regions by the AGN is lower by a factor 4--5 (Heckman et al. 
\cite{heckman2}). No definitive explanation, within the ``unified models'' 
scenario, has yet been given for such a difference (see also Sect. \ref{impli}), but it is 
an observational effect that we have taken into consideration in planning the
experiment. Moreover there are several ISO programs in the Guaranteed Time 
allocation aimed at observing different samples of quasars, including a number 
of CSS/GPS, from which the relevant information could be retrieved.

The project of which we present the results is the merging  of two independent 
programs (P.I.s C.F. and H.F.) which both got observing time in the first ISO 
Call for Proposals. 
A  representative sample of CSS/GPS  radio galaxies, unbiased with respect to 
FIR emission, was selected from various samples of CSS/GPS radio sources.
A sample of extended radio galaxies, matched in 
redshift and radio luminosity, was also selected  for observations at the 
same wavelengths, in order to determine the level of MFIR emission in 
extended radio galaxies as compared to CSS/GPS galaxies.

\smallskip

The layout of the paper is the following.

\noindent
Sect. \ref{sampl} describes the selection of the two samples. 

\noindent
Sects. \ref{obs} and \ref{datar} describe the ISO observations and the data 
reduction. In these sections we mention only what is relevant to the paper.
For more details we refer the reader to the specific ISO and PIA\footnote
{PIA is a joint development by the ESA Astrophysics Division and the ISOPHOT 
Consortium led by the Max Planck Institute for Astronomy (MPIA), Heidelberg. 
Contributing ISOPHOT Consortium institutes are DIAS, RAL, AIP, MPIK, and 
MPIA.} ({\it Iso{\bf  P}hot {\bf  I}nteractive {\bf  A}nalysis}) literature.

\noindent
Sects. \ref{result} and \ref{disc} present the results and discuss them.

\noindent
Conclusions are summarized in bf Sect. \ref{concl}

\section{The sample}
\label{sampl}

Eighteen CSS/GPS radio galaxies in the redshift range $0.2 \leq z \leq  0.8$ 
and with radio size $<  10$ kpc were originally selected. They are mostly
from the 3CR catalogue at 178 MHz (Laing et al. \cite{laing}) and from the 
Peacock \& Wall (\cite{peacock}) (PW) catalogue of radio sources at 2.7 GHz, 
(see Fanti et al.
\cite{fanti2}), but some have been selected also from other lists provided the 
selection was unbiased with respect to IR emission (see Table \ref{sample}).

Of these radio sources, 3C318 has recently been discovered to be a quasar
at a redshift higher than originally believed  (Willott et al. \cite{willott})
and therefore it has to be dropped from the sample, which counts then 17 
CSS/GPS galaxies. However, for completeness,  we have included in Table
\ref{relia} also the data we obtained on this source.

A sample of 16 3CR radio galaxies, with radio sizes $>$ 20 kpc and spanning
a similar range in redshift and radio luminosity, was selected  for comparison 
purposes.
 
The two samples are presented in Tables \ref{sample} \& \ref{sampler}.

Four additional GPSs, at redshifts $< 0.2$ were also selected for
observations (see Sect. \ref{comm}), although they are not going to be used
in the discussion. They also are listed in Table \ref{sample}, separate
from the representative sample.

\begin{table*}
\begin{center}
\caption{Control Sample }
\begin{tabular}{|lrrr|rrrr|} \hline
      &       &                &   &\multicolumn{4}{|c|}{nominal
observing time}\\
Source& z\quad& Log P$_{2.7{\rm GHz}}$& LS& 60$\mu$m& 90$\mu$m& 174$\mu$m&
200$\mu$m\\
& & (W/Hz)\qquad& kpc& sec\quad&sec\quad &sec\quad &sec\quad \\ \hline

\object{3C16}  &   0.41 &   26.31 &       135 & 64 & 32 & 32 & 64\\ 
\object{3C19}  &   0.48 &   26.77 &        20 &  --- &  --- &  --- & 64\\
\object{3C34}  &   0.69 &   26.67 &       174 &  --- &  --- &  --- & 64\\
\object{3C42}  &   0.40 &   26.51 &        85 &  --- &  --- &  --- & 64\\
\object{3C46}  &   0.44 &   26.18 &       533 &256 & 64 & 32 & 64\\
\object{3C79}  &   0.26 &   26.32 &       212 &196 & 32 & 128& 128\\
\object{3C274.1}&   0.42 &   26.52 &       470 & 64 & 32 & 32 & 64\\
\object{3C277.2}&   0.77 &   26.92 &       214 & 64 & 32 & 64 & 128\\
\object{3C284}  &   0.24 &   25.87 &       409 & 64 & 32 & 32 & 32\\
\object{3C295}  &   0.46 &   26.51 &        20 &  --- &  --- &  --- & 32\\
\object{3C299}  &   0.37 &   26.44 &        33 &  --- &  --- &  --- & 32\\
\object{3C330}  &   0.55 &   27.18 &       192 &  --- &  --- &  --- & 32\\
\object{3C337}  &   0.64 &   26.96 &       153 & 64 & 32 & 32 & 32\\
\object{3C401}  &   0.20 &   26.12 &        40 &196 & 64 & 32 & 64\\
\object{3C441}  &   0.71 &   27.03 &       117 & 64 & 32 & 32 & 128\\
\object{3C459}  &   0.22 &   26.18 &        20 & 64 & 32 & 64 & 64\\ \hline\hline
median &   0.43 &   26.51 &       150 &    &    &    &   \\ \hline \hline
\end{tabular}
\label{sampler}
\end{center}
\end{table*}

\section {Observations}
\label{obs}

The sources were observed with ISOPHOT (Lemke et al. \cite{lemke}), sub--instruments 
C100 \& C200, at the wavelengths of 60, 90, 174 \& 200 $\mu$m. These
wavelengths were chosen in order to cover a spectral range as broad as
possible with the best detectors. Actually
some sources, mainly in the control sample,  were observed in other programmes,
including Guaranteed Time Observations, requiring us to observe
them only at the wavelengths not planned in the other experiments (see Tables
\ref{sample} \& \ref{sampler}, where the wavelengths missing in our programme are
marked by a ``---'').

The C100 detector consists of a 3$\times$3 matrix of pixels
$43.5''\times43.5''$ in size  and C200 of a 2$\times$2 matrix of
$89.4''\times89.4''$ pixels. 
Each pixel is an independent detector, which requires its own calibration.

The diameter of the Airy disks of the Point Spread Functions (PSF) are
given in Table \ref{tfpsf}. By comparison with the pixel sizes, it is
clear that for C100 the Airy disk matches approximately one pixel size,
while for C200 the PSF covers most of the matrix. Therefore detected sources
will be visible mostly on the central pixel \#5 with C100, while with C200
the four pixels should give approximately the same values (within the noise).
Table \ref{tfpsf} gives also the fraction, $f_{\rm PSF}$, of light falling
onto the central pixel (\# 5) for C100 and on the four pixels for C200. 

Observations were made in {\it
chopper} rectangular mode, with chopper throw 180'', in order to have, 
every few seconds ({\it chopper plateau}), an ON--source and an 
OFF--source measure of the same time length. The  OFF data then have to be 
subtracted from the ON data, in order to extract the source signal. In 
this observing mode, the telescope points halfway between the source and the
background positions, and a small mirror switches between the two sky
positions. This introduces a vignetting error which is different for ON and OFF
positions (see Sect. \ref{datar}).

We applied for and obtained time 
also in the second ``Call for Proposals", mainly for the purpose of increasing 
the observing time (originally planned on the basis of the ground estimates of 
the instrument performances). Therefore some of the sources were observed 
twice. 

The total ON--source time ranges from 32 to 256 sec, although the glitch 
removal and other options of the data reduction (Sect. \ref{datar}) have 
shortened the nominal observing time. In Tables \ref{sample} \& \ref{sampler}
however, we give the total ``nominal'' ON--source time; for the sources 
observed twice we give the total.

\section {Data reduction}
\label{datar}

\begin{table}
\begin{center}
\caption {Airy disk size, $d_{\rm Airy}$ and $f_{\rm PSF}$ for the filters used in the 
observations}
\label{tfpsf}
\begin{tabular}{|l|r|r|l|r|r|} \hline
filter  &$d_{\rm Airy}$& $f_{\rm PSF}$& filter &$d_{\rm Airy}$& $f_{\rm PSF}$ \\
$\mu$m    &['']      &          & $\mu$m   &['']      &          \\\hline
{\bf C100}&          &          &{\bf C200}&          &           \\
60        &50.3      & 0.69     & 174      &134       &0.707$^a$       \\
90        &75.5      & 0.61     & 200      &168       &0.794$^a$      \\ \hline
\end{tabular}

a) Laureijs (1998) -- private communication
\end{center}
\end{table}

The data analysis was carried on using mainly the PIA V7.2 software (Gabriel
et al. \cite{gabriel}). Some further software has been kindly made available to us by 
Dr. M. Haas from MPIA (Heidelberg) or has been written by ourselves.

PIA removes glitches due to cosmic rays, subtracts the dark current, 
corrects for drifts, derives the signals from the {\it ramps}\footnote{The 
signal is read from the detector every 1/32 sec and accumulated
until the voltage reaches a saturation limit. At this time a ``destructive"
read--out occurs, which resets the integration. The integral signal between 
two consecutive destructive read--outs is called a ``ramp" and its slope 
gives the signal in  Volt/sec.}, calibrates the data and  makes  
corrections for vignetting.

Special PIA features that we used are:

1) We used the method of ramp subdivision, i.e. we divided each ramp into 
pieces 0.25 sec long ({\it pseudo--ramps}) to improve the accuracy of signal 
determination. 

2) to remove glitches we used the ``two threshold glitch recognition" which is
claimed to work better than the original one. In a number of cases, 
however, glitch ``tails" remain in the data. We wrote a simple IDL program to 
further statistically remove the glitch residuals. The noise is reduced by 
approximately a further 20\%.

3) Due to ``memory effects" the chopper plateaus are not always flat, but in
many cases show upwards/downwards trends in the ON/OFF measurements. In absence
of an appropriate program which fits these trends and extrapolates the 
asymptotic
value, we used the quite crude PIA option which removes the first half of
each chopper plateau, thus cutting in half the observing time.

More detailed remarks on some other of these reduction steps are discussed
here below.

\subsection{Calibration}
\label{calib}
The conversion Volt/sec into Watt is made by  using an internal 
calibration source  which provides the conversion factors for each of
the matrix pixels (or detectors). Then the conversion to  Jy is made by
using the known instrument and filter characteristics. The flux density
scale may still be  $\ltsim$ 20\% uncertain at 60 and 90 $\mu$m (see
Sect. \ref{IRAS}) and perhaps more at 174 and 200 $\mu$m. 

In addition, the matrices of the OFF measurements are not ``flat" after PIA
calibration, the pixel--to--pixel fluctuations being much larger than the
instrumental noise. This is explained as a residual calibration error of the
individual pixels (flat fielding), that we estimate to be up to
$\approx\pm 15$\%. This causes fluctuations from pixel to pixel of several tens of mJy,
essentially due to the background, as the ``true" signal from the sources 
is rather small. These errors, however, 
largely cancel out in the (ON -- OFF) data and we have not attempted any 
correction.

Finally, according to the ISO team, it seems that flux densities 
 have to be increased by 
an amount which depends on the chopper frequency (i.e. the rate at which the
chopper switches between ON and OFF positions)
and on the signal difference between ON and OFF. At
the time of writing this paper this correction has not been firmly established
yet, although it is believed to be small for the C200 data. 
For the C100 data only the strong sources
may be affected by this problem (see also Sect. \ref{IRAS}), while 
for the average detections (see Sect. \ref{ave-det}) we expect the
correction to be negligible. 

\subsection{Vignetting corrections}
\label{dev}

The vignetting corrections applied by PIA have an uncertainty of a few percent.
As they are different in the ON and OFF measures, they do not cancel in 
computing (ON -- OFF), as residual flat--field errors do. 
Given the typical values of the background (Table \ref{r.m.s.}), this 
uncertainty  introduces errors which are not negligible as 
compared to  the source flux densities which are rather weak.  

At 60 and 90 $\mu$m, pixel \# 5, where the source signal is largely
concentrated, is not affected by vignetting and the source flux density 
is safely
derived as explained in Sect. \ref{flux} by dividing the signal on pixel \# 5, 
by the $f_{\rm PSF}$.

At 174 and 200 $\mu$m, where the source signal is spread over the four 
pixels, vignetting errors are a major cause of uncertainty, due to the higher 
background levels, not only as they increase the final noise, but also 
as they introduce systematic effects. Indeed, after application of PIA
vignetting corrections, we got the peculiar result that the (ON -- OFF) flux
densities were, on average, systematically negative, with a clear dependence 
on the background brightness.  This effect is very likely due to residual 
vignetting errors. 

We have therefore tried to estimate such corrections 
independently, from our own data.
For this purpose we used a method suggested by Dr. M. Haas, from MPIA, whose 
application we take the responsibility for.
We assumed that the pixels which are more central with respect to the 
telescope pointing, i.e. \# 2 and \#3 at ON, and  \# 1 and \#4 at OFF
position, are, at the first order, unaffected by vignetting, while the other 
pixels are. 

As the PSF fills essentially all four pixels, they do see the same signal. Any
deviation from this is attributed to vignetting.
Considering only the sources with very strong background ($\gtsim $20 MJy/ster),
assuming that the source signal is negligible as compared to the 
background, and that the background does not change appreciably from ON to
OFF, the four pixels at ON position ought to see approximately the same signal 
as the four pixels at OFF. 
Therefore the ratios of the signal in pixels \# 1 and \# 4 ON (assumed to be 
the only ones affected by vignetting) to the signal of the corresponding 
pixels at OFF position (assumed to be unaffected by vignetting) give the 
vignetting errors for these two pixels. The same procedure is applied to pixels
\# 2 and \#3 at OFF position. Note that flat--fielding errors cancel out in 
taking the ratios of corresponding pixels.

The correction factors deduced from the different sources are consistently 
in the range from 3 \% to 5 \% and were applied to 
all the sources. The estimated residual uncertainty is $\approx$ 1.0 \%.
The application of these corrections eliminates the problem of the systematic 
negative (ON -- OFF) values and of their dependence on the background
brightness.

\subsection{Flux density determination}
\label{flux}
For C100 the source flux density should be derived by fitting a PSF of known 
width to
the (ON -- OFF)  matrix values. In practice our signal is always very weak 
and will only be detectable on pixel \#5. Therefore we have computed the 
source flux density simply by dividing the flux density falling onto pixel \#5 by 
the 
appropriate value of $f_{\rm PSF}$ given in Table \ref{tfpsf}.

For C200 the flux density has been obtained by summing the (ON--OFF) four 
matrix pixels
and dividing the result by the $f_{\rm PSF}$ in Table \ref{tfpsf}.

In the following we shall therefore use as units {\it mJy/pixel} to refer to 
the flux density falling onto {\it one single} pixel and {\it mJy} to indicate
source flux density.

\subsection{Estimate of statistical flux density errors}
\label{errors}
Besides the calibration and vignetting errors, additional uncertainty in the 
flux density measurements is caused by: {\it a)} instrumental
noise (or simply {\it noise}, $\sigma_{\rm n}$), which includes all noise 
contributions along the signal chain (i.e. photon noise due to source and
background, plus detector noise), {\it b)} confusion due to foreground 
galactic {\it cirrus} 
( $\sigma_{\rm cc}$) and {\it c)} confusion due to extragalactic background 
($\sigma_{\rm egc}$). Typical values are reported in Table \ref{r.m.s.}.

\medskip
In the next paragraphs we discuss how they have been evaluated in the paper.

\medskip\noindent
{\it a) Instrumental noise, $\sigma_{\rm n}$ }  

The nominal instrumental noise {\it per pixel} is given by the equation:
\begin{equation}
\label{noise}
\sigma'_{\rm n} =\sqrt{NEP^2_{\rm source}+2NEP^2_{\rm bck}+2NEP^2_{\rm rec}}
\end{equation}

\noindent
where the $NEP$'s are the {\bf N}oise {\bf E}quivalent {\bf P}ower of source,
background and receiver, to be computed using parameters given in the ISO
manuals and which depend on the ON--source time. The factors of 2 in Eq. (\ref{noise})
derive from the fact that the signal is obtained from (ON -- OFF) measurements.
>From the values of the parameters necessary to compute the $NEP$'s it is
clear that, except for very high backgrounds ($\gtsim 100$ MJy/ster) or
for very strong sources ($\gtsim$ 50--100 Jy), the contribution of the 
detector to the nominal noise is expected to dominate.

Because the resulting performance of the instrument was worse than expected,
we adopted a pragmatic method and estimated $\sigma'_{\rm n}$ from the data
themselves, instead 
of using the parameters given in the manuals.

To estimate $\sigma'_{\rm n}$ directly from the data, within each observation 
we used the chopper sequence (i.e.
the sequence of ON--OFF pairs of each source) to determine the statistics of 
the (ON -- OFF) 
values. For C100 we used only pixel \# 5, thus avoiding the residual
flat--field (Sect. \ref{calib}) and vignetting (Sect. \ref{dev}) problems.
For C200 we used the four pixels together, after application of vignetting
corrections, and ignoring the residual flat--field corrections.

The average (ON -- OFF) gives the flux density within pixel \# 5 for C100 and the
flux density within any of the four pixels for C200. The r.m.s. distribution of 
the
ON--OFF pairs gives the ``actual instrumental noise" $\sigma'_{\rm n}$
(in mJy/pixel) of {\it that one} observation. The software for this analysis
is due to M.Haas (private communication).
                                                                     
We find that $\sigma'_{\rm n}$  can change significantly from source to source,
even for similar backgrounds and observing times. The {\it median} values,
however, scale rather well with the inverse of the square root of the
ON--source time. We do not find any
difference as a
function of the background brightness, which confirms that the detector noise
is the dominant factor. The noise computed in this way is in
agreement with that from Eq. \ref{noise} when adopting the parameters of the 
ISO manual, except for $\lambda = 200 \mu$m where our estimate is lower. 
This proves, to us, that no significant systematic error is present in our 
data and that glitches have been removed fairly well.

Given the good behavior of the instrumental (mostly detector's) noise with 
$1/\sqrt{t}$, we give in
Table \ref{r.m.s.}, for each observing wavelength, one single value for
$\sigma'_{\rm n}$, corresponding to an integration time of 64 sec. 
Note that with C200,
since the four pixels are independent detectors, the flux density error due to the
noise will be computed as $\sigma_{\rm n}=2\times \sigma'_{\rm n}/f_{\rm PSF}$.

\medskip\noindent
{\it b) Cirrus confusion noise, $\sigma_{\rm {cc}}$ }

This term has been evaluated by Gautier et al. (\cite{gautier}) 
who provide useful tables to 
compute $\sigma_{\rm {cc}}$ on any angular scales
(3 arcmin, in our case, the value of the chopper throw) and also show that
cirrus confusion depends strongly on the galactic background brightness, $B$, 
according to  $\sigma_{\rm cc}\propto B^{1.5\pm 0.2}$.

Since  our  measurements include a contribution from the celestial 
fore/back--grounds and, 
at least at 60 and 90 $\mu m$, a strong contribution from
zodiacal light, in order to  estimate, $\sigma'_{\rm {cc}}$ (within a pixel), at 
all four wavelengths, for both C100 and C200 we used  IRSKY software by IPAC
which applies a model to remove the zodiacal light, providing 
also the values for a pure galactic foreground.
This estimate of $\sigma'_{\rm {cc}}$ is accurate to within a factor of two. 

Note that, for $\lambda > 100 \mu$m, the foreground provided by IRSKY 
is  extrapolated from the IRAS shorter wavelengths, using the
dust emission model by Desert et al. (\cite{desert}). This increases the uncertainty on
$\sigma'_{\rm {cc}}$. For consistency and to be conservative we generally used the 
IRAS 
background, which is, on average, higher than ISO's, except in
those cases in which, by comparison with the other satellite values, the
extrapolation from IRAS was obviously wrong. In those cases we used the ISO 
background.

We have also compared IRSKY values of $\sigma'_{\rm {cc}}$ with those reported in 
ISO manuals:
while at 60 and 90 $\mu$m the two estimates do agree, at $\lambda=174\mu$m
the ISO manual estimate is over one order of magnitude greater than IRSKY's. 
We found no estimate at 200$\mu$m. 

As a reference, we report in Table \ref{r.m.s.} the values of $\sigma'_{\rm
{cc}}$
at each wavelength, from IRSKY, for a foreground brightness typical of our 
observations.

\medskip\noindent
{\it c) Extragalactic confusion noise, $\sigma_{\rm egc}$ } 

This is simply due to the piling up, within a beam, of faint galaxies not 
individually resolved by the instrument. The estimate of the magnitude of
this effect, within
each pixel, is reported in Table \ref{r.m.s.} and is derived by IRSKY, in the
hypothesis of cosmological evolution of faint galaxies. These values are
about a factor of two higher than in the case of a non--evolving model.

Note that with C200, since the PSF covers most of the four pixels, we 
considered the cirrus and extragalactic confusion correlated across the 
matrix and therefore we adopted, as the flux density error, 
$4 \times$(pixel error)/$f_{\rm PSF}$.

\medskip
The total error on source flux density is then computed as:
\begin{equation}
\sigma{\rm _T}=\sqrt{\sigma_{\rm n}^2+\sigma_{\rm egc}^2+\sigma_{\rm {cc}}^2}
\label{sig-t}
\end{equation}
where $\sigma_{\rm n}$ is the {\it individual source} noise, $\sigma_{\rm egc}$ is the 
error due to the
extragalactic confusion (which is the same in all fields at a given wavelength) 
and $\sigma_{\rm {cc}}$ the
cirrus confusion noise appropriate for the specific source background
(zodiacal light subtracted). In Table \ref{r.m.s.} total flux density errors, 
appropriate to the parameters in the Table, are also given as an example.

As it can be seen from Table \ref{r.m.s.}, the extragalactic source and
cirrus confusion ($\sigma_{\rm egc} $
and $\sigma_{\rm {cc}}$)  may be not negligible as compared to the instrumental
noise and may limit the instrument sensitivity considerably or lead to spurious
detections, as it will be further discussed in Sect. \ref{detect} .

As an internal test, we used those sources observed twice (with different
backgrounds) to check the goodness of error evaluation.

At each wavelength, we compared the difference $\Delta S$ of the two measured 
flux densities  with the combined total error ($\sigma^c_{\rm T}$) 
obtained by adding in quadrature the $\sigma_{\rm T}$ of the two observations. 
If errors are estimated properly, the ratio
$\Delta S/\sigma^c_{\rm T}$, computed for all the available pairs, should have
zero average and standard deviation =1.

In spite of the very poor statistics (8 sources observed twice with C100 
and 4 with C200) the agreement is satisfactory.

\begin{table}
\begin{center}
\caption{Noise, confusion and total flux density errors }

\begin{tabular}{|r|rrlr|r|} \hline
$\lambda$ & $\sigma'_{\rm n}$(64 sec)& $\sigma'_{\rm {cc}}~~~$& bckg&
$\sigma'_{\rm egc}~~~$&
   $\sigma_{\rm T}$ \\
$\mu$m &   mJy/pix  & mJy/pix      &MJy/sr& mJy/pix& mJy\\ \hline
  60 &      27~~~~ &       9~~~~~& ( 2)&  2~~~~~&                  42\\
  90 &      12~~~~ &       5~~~~~& ( 2)&  4~~~~~&                  22\\
 174 &      23~~~~ &       8~~~~~& (10)&  22~~~~~&                148\\
 200 &      61~~~~ &       6~~~~~& (20)&  23~~~~~&                195\\  \hline
\end{tabular}
\label{r.m.s.}

\end{center}
Values in parentheses are the background brightness at which $\sigma'_{\rm
{cc}}$
is computed
\end{table}

\section{Results}
\label{result}

\subsection{Individual Detections}
\label{detect}

About  half of the observed radio galaxies, both in the representative sample 
of CSS/GPS and  in the 
comparison sample, have $S> 3 \times  \sigma_{\rm n}$, at least at one wavelength.
These could be considered ``formal'' detections, in the sense that we got
{\it some signal} out of the instrumental
noise. This {\it does not mean}, however, that we have {\it detected our 
target sources}, since, as seen in Table \ref{relia}, extragalactic and/or
cirrus confusion could cause spurious detections: we might just
have detected a cirrus filament or a fluctuation in the confusion. 

We have checked the reality of these ``formal" detections by looking for
``negative" detections. We have only three such cases (see
Tables \ref{relia} and \ref{undet}) suggesting that, after all, in the
``formal'' detections some signal from the target is
present. This is why we have kept them in Table \ref{relia}. The test
for ``claiming" a detection, however, is made by comparing $S$ with
$\sigma_{\rm T}$  (Eq. \ref{sig-t}).

We considered {\bf detected} the sources for which $S\geq 5 \times
\sigma_{\rm T}$ 
at least at one wavelength (value in {\bf bold} in Table \ref{relia}) and 
{\it possibly detected} those for which $ 3 \times \sigma_{\rm T}\leq S < 5 \times 
\sigma_{\rm T}$ (value in {\it italic} in Table \ref{relia}).

Table \ref{relia} gives the detected, possibly detected and ``formally'' 
detected sources, listing the flux densities  at all four wavelengths 
(significant or not) and related errors. The sources detected at least at one 
wavelength are marked with $\ast\ast$; those possibly detected are marked 
with $\ast$. In each column we give $S, \sigma_{\rm n}$ and $ \sigma_{\rm T}$ in mJy (1 $\sigma$ level).

\begin{table*}
\begin{center}
\caption{Sources detected at different levels of reliability (mJy)}
\begin{tabular}{|lr|rrr|rrr|rrr|rrr|} \hline
&Source&\multicolumn{3}{|c|}{$\lambda=60$ $\mu$m}&
\multicolumn{3}{c|}{$\lambda=90$ $\mu$m}&
\multicolumn{3}{c|}{$\lambda=174$ $\mu$m}&
\multicolumn{3}{c|}{$\lambda=200$ $\mu$m}\\ \hline
&&S&$\sigma_{\rm n}$&$\sigma_{\rm T}$&S&$\sigma_{\rm n}$&$\sigma_{\rm T}$&S&$\sigma_{\rm
n}$&$\sigma_{\rm T}$&
S&$\sigma_{\rm n}$&$\sigma_{\rm T}$\\ \hline
$\ast$         &3C49  & 42  & 14& 15& 16 &  17& 23& 335 &  52& 146& {\it 623}&102&
155\\
 &1244+49& 22  & 18& 21& 36 & 16& 17& 187 &  48& 133&  55&134& 177\\
  $\ast$&1323+32 & 1   &16 &17 & {\it 45} &  11& 13& 187 &  36& 130& 306&108& 
157\\
  $\ast$&3C303.1 &  --2 &18 &23 & 45 & 17 &19 &{\it 543}  &117 &172 &368 &248& 
274\\
  $\ast$&1622+663& 77  &30 &33 &  {\it 43}&   8& 12& --85 &  73& 145&  81&  89& 
146\\
 $\ast\ast$&1819+39 & {\it 91}  &19 &21 & {\bf 147}&   6& 12&   33 &  53& 136&   
8&132& 176\\
    &1829+29 &99  &40 &51 & 129& 36& 80&--741 &  44& 313&--348&107&
171\\
  $\ast$&2352+49 & 8   &59 &61 & 66 & 27 &51 &204  & 47 &343 &{\it 655} &130& 
202\\ \hline
  $\ast$&3C318   & {\it 118} &35 &36 & {\it 104}&  21& 24& 277 &  60& 150& 239&  
68& 135\\
    &      &     &     &     &    &      &     &     &      &      &    &&\\
 $\ast\ast$&1345+125 & {\bf1470}& 45& 46&{\bf 1171}&  23& 24& {\bf 1273}& 
119&172&{\bf 1471}&287& 310\\
 $\ast\ast$&1718--64 & {\bf 499}  &27 &33 &  {\bf 602}&  13& 44&{\bf 2331} &  
41&396&{\bf 2216}& 77& 184\\
&1946+70 &56   &49 &51 &  10&  11& 41& --28 &  52& 334& 433&105& 184\\ \hline

&3C16& --77 & 33& 44& --16&  25& 33& 181 &  47& 164& 393&156& 197\\
 $\ast$&3C34 &     &     &     &    &      &     &     &      &      &{\it 781} 
&149& 193\\
 &3C46&  11 & 17& 17&  32& 10 &13 & 283 & 115& 177& 212&162& 200\\
 $\ast$&3C79 & 66  &24 &29 & 37 &14  &37 &378  & 29 &326 &{\it 625} &103& 181\\
  $\ast$&3C277.2 & {\it 87}  &25 &27 &--18 & 13 &16 &289  & 42 &133 &207
&152& 191\\
 $\ast$&3C284& {\it 179} & 47& 47&  18& 29 &30 & 51  &150 &197 & 86 &180& 214\\
 &3C401& 2   &14 &17 &  32& 18 &22 &238  & 55 &158 &469 &213& 244\\
    &3C441 & 5   &28 &30 & --7 &19  &29 & 11  & 80 &246 & 453 &120& 178\\
 $\ast\ast$ & 3C459 &{\bf 698}  &44 &51 & {\bf 747}& 45 &48 &{\it 668}  & 43 
&157&{\bf 1164}& 157& 196\\ \hline
\end{tabular}
\label{relia}                                                                          
\end{center}
flux densities  in {\bf bold} characters are those for which $S>5 \times 
\sigma_{\rm T}$;

flux densities  in {\it italic} characters are those for which $ 3 \times 
\sigma_{\rm T}\leq S\leq 5 \times \sigma_{\rm T}$
\end{table*}

In Table \ref{relia} only two sources from the representative (1819+39) and 
control (3C459) samples are clearly detected at least at one wavelength. 
Nine more sources have flux densities which
exceed $3~ \sigma_{\rm T}$, but are  $< 5 ~ \sigma_{\rm T}$, and are therefore possible
detections. 

The quasar 3C318 is detected at a $\geq 3 \sigma$ level at both 60 and 90 
$\mu$m.

We also detect clearly at all  four wavelengths two of the four nearby GPS
(1345+125 and 1718--64).

In Table \ref{undet} we list the undetected sources with their $S$, 
$\sigma_{\rm n}$ and $ \sigma_{\rm T}$, in mJy.

\begin{table*}
\begin{center}
\caption{Undetected Sources (mJy)}
\begin{tabular}{|r|rrr|rrr|rrr|rrr|} \hline
Source&\multicolumn{3}{|c|}{$\lambda=60$ $\mu$m}&
\multicolumn{3}{c|}{$\lambda=90$ $\mu$m}&
\multicolumn{3}{c|}{$\lambda=174$ $\mu$m}&
\multicolumn{3}{c|}{$\lambda=200$ $\mu$m}\\ \hline
&S&$\sigma_{\rm n}$&$\sigma_{\rm T}$&S&$\sigma_{\rm n}$&$\sigma_{\rm
T}$&S&$\sigma_{\rm n}$&$\sigma_{\rm T}$&
S&$\sigma_{\rm n}$&$\sigma_{\rm T}$\\ \hline

  0108+389&   9 &19& 19&   5&  11& 14&   45&  40&149& 242&106& 158\\
3C67 &  --2 &12& 18&  28&  13& 33&     &      &     &    &&     \\
0404+76 &   7 &37& 39& --14&  18& 33& --645& 95 &286&--217&170& 220\\
3C268.3 &  17 &22& 24&   2&  15& 17&   34& 57 &137& 151&224& 252\\
1358+624&  54 &30& 31&  22&  29& 30&  124&168 &210& 222&213& 243\\
1607+26 &  71 &60& 62&  83& 43 &45 & --11 &122 &181& 302&223& 252\\
3C343.1 &     &    &     &    &      &     &  --175& 84 &152&--151&102& 154\\
1819+670&  23 &13& 16& --26&  27& 29&   23& 86 &157& --55&199& 231\\
2342+82 &  --3 &30& 33&  32&  14& 36&   --68& 81 &329&   0&219& 265\\ \hline
1934--63 &     &    &     &    &      &     &      &      &     &  40& 93&
154\\ \hline
 3C19&     &     &     &    &      &     &     &      &      &630 &229& 256\\
3C42&     &     &     &    &      &     &     &      &      &348 &132 &179\\
3C274.1 &  53 &49& 54& --18&  32&33 &   6 & 71 &144&--237&128& 173\\
3C295 &     &    &     &    &      &     &     &      &     &--111&102& 154\\
3C299 &     &    &     &    &      &     &     &      &     & 358&148& 188\\
3C330 &     &    &     &    &      &     &     &      &     & 191&246& 272\\ 
3C337& --21 & 33& 34&  43&  17& 18&--141 &  58& 138& 353&192& 224 \\  \hline
\end{tabular}
\label{undet}
\end{center}
\end{table*}

\subsection{Comparison with IRAS data}
\label{IRAS}

As stated in the {\it IRAS Explanatory Supplement} (1985) the sensitivity
limits (3$\sigma$) of the all--sky survey for point sources (PSC) are quite high
($\sim$0.5 Jy at 60 $\mu$m and 1.5 Jy at 100$\mu$m) but can be
sensibly reduced by co--adding several scans, as done, for instance by
Heckman et al. (\cite{heckman2}; \cite{heckman3}), who 
obtained 3$\sigma$ detections on individual
objects of $\approx$100 and $\approx$400 mJy at 60 $\mu$m and 100$\mu$m
respectively. The most sensitive IRAS observations were obtained by
pointed observations (see e.g. Neugebauer et al. \cite{neugebauer}; Golombek 
et al. \cite{golombek}).
The resulting noise varies from field to field and is typically in the
range of 50--100 mJy ($3\sigma$) at 60$\mu$m and 100--300 mJy ($3\sigma$) at 
100$\mu$m. These values  have to be compared with our typical 3$\sigma$
errors, given in Tables \ref{relia} and \ref{undet}, of 
$\approx$90 mJy at 60 $\mu$m and $\approx$75 mJy at 90$\mu$m (median values).
Therefore, while at $\lambda=60\mu$m ISO is not much more sensitive 
than IRAS, at $\lambda=90\mu$m the situation is improved.

We searched the literature to see whether any of the sources in our sample
has been detected by IRAS at any wavelength. The galaxies from our 
lists detected by IRAS at 60 and/or 100 $\mu$m are reported in Table \ref{iras}.

\begin{table}
\begin{center}
\caption{Available IRAS flux densities (mJy)}
\begin{tabular}{|r|r|r|l|} \hline
Source&60$\mu$m&100$\mu$m&ref.\\ \hline
3C79& 179$\pm$9& $< $330& 1\\
    &      &       &  \\
1345+125& 2098& 1738&2\\
&      &       &  \\
3C318 & 170$\pm$26& 300$\pm$70& 1\\
&      &       &  \\
3C459&683& 708&2 \\
&      &       &  \\\hline
\end{tabular}
\label{iras}
\end{center}

ref:  1) Heckman et al. (\cite{heckman2}); 2) Golombek et al. (\cite{golombek})
\end{table}

On the strongest source (1345+125) about 30\%
of the IRAS flux density is missing in our observations. This is the case where 
chopper frequency corrections, that we have not applied  (Sect. \ref{flux}), 
are expected to be larger.  For 3C459, which is not as strong, the agreement with 
the IRAS data is rather good. We are therefore confident that the
missing correction is definitely minor for the fainter sources and
negligible for the median flux densities  that we report.
For the fainter sources, we note that
3C79 is possibly detected in our observations only at wavelengths longer than
100$\mu$m and that 3C318 at $90 ~ \mu$m is fainter than in IRAS, although the 
two flux densities are still within $\sim$ 2.5 of the combined errors.

Conversely 3C284 possibly detected (3.8 $\sigma$ level) at 60 $\mu$m in our ISO 
observations, 
has only an upper limit in IRAS (Impey \& Gregorini 
\cite{impey}) not too different from our measurement. The remaining sources
detected (1819+39) or possibly detected at 60 and/or 90 $\mu$m by ISO, are
weaker than the IRAS upper limits (Heckman et al. \cite{heckman3}).

\subsection{Comments on some individual sources}
\label{comm}

1345+125.
 
\noindent
This radio source is identified with a galaxy with ``two nuclei",
of which the western one has an optical spectrum consistent with
Narrow Line radio galaxies and Seyfert type 
II (Gilmore \& Shaw \cite{gilmore}). There is 
some confusion in the literature about which of the two hosts the radio source.
Gilmore \& Shaw (\cite{gilmore}) and  Baum et al. (\cite{baum}) indicated the eastern 
(non Seyfert nucleus). More recently, however, Capetti (1998, private 
communication), Stanghellini et al. (\cite{stanghellini2}) and
Evans et al. (\cite{evans}) presented convincing  evidence for the association of 
the GPS source with the western Seyfert nucleus. This double nucleus galaxy
is considered a merger which has triggered the radio activity (e.g.
Heckman et al. \cite{heckman}; Sanders et al. \cite{sanders}; Surace et al. \cite{surace}).
Although the optical spectrum is characterized by Narrow Lines, broad 
Pa$\alpha$ (FWHM $\sim 2600$ km s$^{-1}$) is seen in the 
near-IR (Veilleux et al.  \cite{veilleux}).
In addition, Hurt et al. (\cite{hurt}) report
HST/FOC ultraviolet observations which show extended polarized UV
continuum light. Both the UV polarization and the broad Pa$\alpha$
indicate that 1345+125 has a hidden (but not too deeply hidden) quasar.
This is the first GPS radio galaxy detected in X-rays
($L_{\rm X(2-10 Kev}) \approx 2 \times
10^{43}$erg s$^{-1}$, O'Dea et al. \cite{odea4}).

We estimate a temperature of $\sim 30$ K and a dust mass of
$\sim 3.5\times 10^8$ M$_\odot$ for the component which makes
the dominant contribution to our ISO data (Table~\ref{bright-s}).
Although our ISO observations are not sensitive to
gas at temperatures below about 25 K, the CO observations are
sensitive to gas with kinetic temperatures to within a factor of a 
few of the microwave background
temperature (e.g. O'Dea et al. \cite{odea}; Allen et al. \cite{allen}).
For a standard Galactic gas/dust ratio of 100  this corresponds to a mass
of molecular gas $\sim 3.5\times 10^{10}$ M$_\odot$ which is
in good agreement with the estimate derived from CO observations
($\sim 3\times 10^{10}$ M$_\odot$, Mirabel et al. \cite{mirabel}; 
Evans et al. \cite{evans}). 
This agreement between the ISO and CO estimates
of the molecular gas mass may be fortuitous since the gas-to-dust
ratio may vary between galaxies and in general is not well determined
 (e.g. Goudfrooij et al. \cite{goudfrooij}; Falco et al. \cite{falco}).

\bigskip

\noindent
3C 318  

\noindent
This object was classified as an N--galaxy at a redshift 0.75 by Spinrad 
\& Smith (\cite{spinrad}) and considered to be a Narrow
Line (NL) radio galaxy by Hes et al. \cite{hes}. Recently Willott et al.
(\cite{willott}), from UKIRT spectroscopic observations, detected broad
H$_{\alpha}$ and revised the redshift at 1.574. The object is now classified
as a quasar. It is a strong X--ray emitter (Taylor et al., \cite{taylor}) with
a luminosity, updated for the new redshift, of $L_{\rm X(0.5-4.5 Kev})
\approx 8  \times 10^{44}$erg s$^{-1}$, 
\bigskip

\noindent
1622+663:

\noindent
This object  has a broad a H$_{\alpha}$ line. It can be classified as a
Seyfert 1.5 (Snellen et al. \cite{snellen2}). Its broad--line region seems to
be heavily obscured. It is marginally detected at 90 $\mu$m.

\bigskip

\noindent
1718--649

\noindent
The source is identified with the 12.6 blue magnitude galaxy NGC 6328 
(Savage \cite{savage}), whose nucleus has a LINER type optical spectrum (Fosbury
et al. \cite{fosbury}). The galaxy is quite peculiar, having the appearance of a 
high luminosity elliptical with faint outer spiral structure and containing
a large amount of atomic hydrogen (Veron-Cetty et al. \cite{veron}). The object is regarded as a merger of
two galaxies, at least one of which is a gas-rich spiral, in the process
of forming an elliptical. 
A very strong source (a star?) is listed in the IRAS PSC, at 3.7
arcmin from 1718--649. Inspection of IRAS maps at 60 and 100 $\mu$m shows a 
fainter extension of this bright source to the North, at the position of  
1718--649.

The dominant component of dust contributing to our ISO detection
has an estimated temperature of 20 K and a mass of 
 $\sim 0.1\times 10^{8}$ M$_\odot$ (Table~\ref{bright-s}) 
which would correspond to a
gas mass of $\sim 10^9$ M$_\odot$. This is at least an order of
magnitude less than the mass of HI ($\sim 3.1\times 10^{10}$ 
M$_\odot$,  Veron-Cetty et al. \cite{veron}).  The HI is distributed
over the entire galaxy and much of it lies in a partial ring with 
a diameter of $95"$ (19 kpc). Possibly, much of the dust in this 
object lies at large distances from the nucleus and is relatively 
cold, thus escaping detection in our ISO observations. CO observations
would be useful.

\bigskip

\noindent
1819+39

\noindent
Its redshift has been re--measured by Vermeulen et al. (\cite{vermeulen}) who
also show that this object has broad H$_\beta$. Thus, this object
appears to be Broad Line radio galaxy.

\bigskip

\noindent
1934--638

\noindent
The radio source is associated with the brighter galaxy of a pair of compact
galaxies ($\approx$ 8 kpc separation) sharing a common envelope (Jauncey
et al. \cite{jauncey}). Fosbury et al. (\cite{fosbury2}) suggest larger amounts of gas and dust
than generally  found in typical Narrow Line radio galaxies.
Tadhunter et al. (\cite{tadhunter}) found presence of polarized light,  suggesting
an anomalous environment (see also Morganti et al. \cite{morganti}). It is
undetected at the only observed wavelength (200 $\mu$m).

\bigskip

\noindent
1946+70

\noindent
This is one of the three GPS sources from the sample of Snellen et al. (\cite{snellen}),
and it is much weaker than those from the 3CR and PW samples. It is
``formally'' detected at 200$\mu$m.

\bigskip

\noindent
3C459

\noindent
The optical object has been classified as N--galaxy by Spinrad et al. 
(\cite{spinrad2}). It 
was studied in detail at radio and optical frequencies by Ulvestad
(\cite{ulvestad}), who 
concluded that this galaxy may have undergone a large amount of fairly recent 
star formation. Hes et al. (\cite{hes}) classified it as Narrow Line 
radio galaxy and suggest 
that a fraction of the FIR emission may be due to heating of the dust by the 
moderately young blue stars.

\subsection{Average Detections}
\label{ave-det}

We have looked for ``statistical'' emission from both the representative
CSS/GPS sample (i.e. for $0.2 \leq z \leq 0.8$) and the control sample. We have 
computed both the mean and the median flux density, although in the following 
we use the second one, which is considered to be
more robust since it is less affected by extreme flux density values.
In computing the mean flux density we excluded 3C459 for C100, since its large
detected flux density would affect the results too much (the source was used,
instead, in computing the median). 

Values are given in Table 
\ref{stat}. The uncertainty on the mean and median are derived from the 
flux density distribution itself. Note that, as the error distribution 
of the median is not Gaussian, we choose to give the 95\% confidence
uncertainty, equivalent to a 2 $\sigma$ limit for a Gaussian distribution.
To make the comparison easier we then give 2$\sigma$ errors also on the mean.

At 60 and 90 $\mu$m there is clearly a 3 $\sigma$ detection for the CSS/GPS 
sample. In the comparison sample, where the 
statistics is poorer, as we do not yet have data for $\approx ~1/3$ of the
objects (see Table \ref{sampler}), both mean and median flux densities,  
compared with the errors, do not indicate a significant detection.
They are, however, largely consistent with the CSS/GPS results. 
This is reinforced by a comparison done with Heckman et al. (\cite{heckman2};
\cite{heckman3}). These authors give co--added flux densities for a sample of 
radio galaxies.
In their ``near sub--sample'' ($0.3 \ltsim z \ltsim 0.85$, median $z=0.48$), 
composed of 5 CSS/GPS and 36 large size sources, and therefore 
similar to our comparison sample, 
the median flux densities (S$_{60} \approx 25\pm$5 mJy, S$_{100} \approx 
40\pm$12)  are consistent with ours, but have smaller errors, due 
to the much larger number of sources in their sample.

At 174$\mu$m and 200$\mu$m  radio galaxies in the control sample seem
stronger FIR emitters, although the difference is not significant at
either wavelength. Averaging the two wavelengths the difference is around
2$\sigma$.

 We have also applied a Kolmogoroff--Smirnov test, which shows that the 
hypothesis that the two galaxy samples are extracted from the same population 
is acceptable at a 99\% confidence level at all wavelengths.

\tabcolsep 1.1mm
\begin{table}
\begin{center}
\caption{Average flux densities for the CSS/GPS and the comparison sample}
\begin{tabular}{|r||rcl|c||rcl|c|} \hline
$\lambda$&\multicolumn{4}{c||}{CSS/GPS}&\multicolumn{4}{c|}{Control}\\
&\multicolumn{3}{c}{mean}&\multicolumn{1}{c||}{median}&
\multicolumn{3}{c}{mean}&\multicolumn{1}{c|}{median}\\
$\mu$m& \multicolumn{4}{c||}{mJy}&\multicolumn{4}{c|}{mJy} \\ \hline
   60&36&$\pm$& 24  &  $20^{+34}_{-13}$& 34 &$\pm$& 48& $32^{+55}_{-30}$\\
      &   &    &     &                  &    &     &   &                 \\
   90&39&$\pm$& 22  &  $34^{+11}_{-18}$&  11 &$\pm$&  18& $25^{+15}_{-40} $\\
   &   &    &     &                  &    &     &   &                 \\
   174&--2&$\pm$& 156  &  $34^{+90}_{-119}$& 198&$\pm$&146& $210^{+90}_{-195}$
   \\
   &   &    &     &                  &    &     &   &                   \\
   200&140&$\pm$& 130 &  $116^{+186}_{-116}$&370&$\pm$& 168& $355^{+120}_{-145}$\\
   &   &    &     &                  &    &     &   &                 \\ \hline   
   \end{tabular}
\label{stat}

errors on both mean and median are 2 $\sigma$ (see text)
\end{center}
\end{table}

Considering also the results from Heckman et al. (\cite{heckman2};
\cite{heckman3}), we estimate that any IR excess for CSS/GPS sources does not
exceed (at a 95\% confidence level) $\sim$ 25 mJy in the range 60--90 $\mu$m
and $\sim$ 100 mJy in the range 170--200 $\mu$m.

Our first conclusion then is that we do not have any statistical evidence 
that the MFIR properties of the galaxies associated with CSS/GPS sources are
much different from those of the galaxies associated with large size radio 
sources (see also Sect. \ref{fir-l}) consistent with the results of
Heckman et al. (\cite{heckman3}). The small and large sources we
observed seem to have 
similar average luminosities at all four wavelengths, and hence a similar 
spectrum between 60 and 200 $\mu$m (observer frame). This will allow us 
to combine the two sets of data in the discussion (Sect. \ref{fir-l}).

\section{Discussion}
\label{disc}

\subsection{FIR tail of synchrotron emission}

We have collected all the flux densities available in the literature at radio
wavelengths for every source in the sample. No evidence is seen for bright flat
spectrum cores at the highest available frequencies (often up to 230 GHz,
1300 $\mu$m).
The spectra can be well fitted either with a power law   or  with a 
model which includes spectral curvature due to synchrotron losses (see Murgia 
et al. \cite{murgia}, for details). For each source, using the fitted spectrum, 
we have extrapolated the flux densities at 60, 90, 174 and 200 $\mu$m. 
The extrapolated average synchrotron flux density is very small 
($\ltsim$ 2 mJy at 200 $\mu$m and less at the other 
wavelengths) and does not contribute significantly to the FIR emission. 
Therefore we attribute the emission we find to genuine thermal radiation.

\subsection{FIR luminosities of the {\rm combined} CSS/GPS and comparison 
sample} 
\label{fir-l}

As discussed in Sect. \ref{ave-det}, our data do not show any
statistically significant evidence for
differences in the FIR luminosities of CSS/GPS and large size radio
galaxies at 60 and 90 $\mu$m, the situation being somewhat more uncertain 
in the 174--200 $\mu$m interval.

Now we combine the two samples together into
a single {\it sample of powerful radio galaxies at intermediate redshift}
to be studied on its own, independent of the type of associated radio source.
In Table \ref{whole},  we report the mean and median FIR flux densities 
of the combined sample and the average FIR luminosities in the 60--90$\mu$m 
and 174--200 $\mu$m ranges.

\begin{table}
\begin{center}
\caption{Average flux densities and FIR luminosities for the whole sample}
\begin{tabular}{|r||rcl|c||r|r|} \hline
$\lambda$&\multicolumn{3}{c}{mean}&\multicolumn{1}{c||}{median}&
L$_{\rm FIR}$/L$_\odot$ &L$_{\rm FIR}$\\
$\mu$m& \multicolumn{4}{c||}{mJy} &&erg s$^{-1}$\\ \hline
      60 & 35 &$\pm$& 22 & 20$^{+34}_{-13}$&&                  \\
         &    &     &    &                 & $6 \times 10^{10}$&
$\approx 2.4\times 10^{44}$\\
      90 & 29 &$\pm$& 16  & 32$^{+15}_{-16}$& &                  \\
         &    &     &    &                 &  &                 \\
   174 & 74 &$\pm$& 114 & 48$^{+139}_{-37}$&  &                  \\
         &    &     &    &                &  $\approx 10^{11}$&
$\approx 5\times 10^{44}$  \\
   200 & 256&$\pm$& 112& 232$^{+123}_{-114}$& &                 \\ 
         &    &     &    &                 &&\\ \hline
   \end{tabular}
\label{whole}

errors on both mean and median are 2 $\sigma$

\end{center}
\end{table}

\bigskip

The FIR luminosities from the median values of the whole sample
are calculated as:
\[L_{\rm FIR} = 4 \pi D_{\rm L}^2  \nu S_{\nu}\]

\noindent
where $S_{\nu}$ is the median flux 
density at the frequency $\nu$ and $D_{\rm L}$ is the
luminosity distance. 

The average $L_{\rm FIR}$ are in the range $(0.6-1.0)\times 10^{11}$ 
$L_\odot$, or (2--5)$\times 10^{44}$ erg s$^{-1}$. The brightest objects, 
1819+39 and 3C459 , which are several times more luminous than average 
(Table \ref{bright-s}), have peculiarities, as commented in Sect. \ref{comm}, 
which may well account for their larger FIR luminosities.

These luminosities are much higher than those found in nearby radio quiet 
elliptical galaxies, which are in the range of up to $\approx 10^8 L_{\odot}$
(Goudfrooij \& de Jong \cite{goudfrooij2}; Bregman et al. \cite{bregman}). 

\begin{figure}
\vspace{7 cm}
\includegraphics{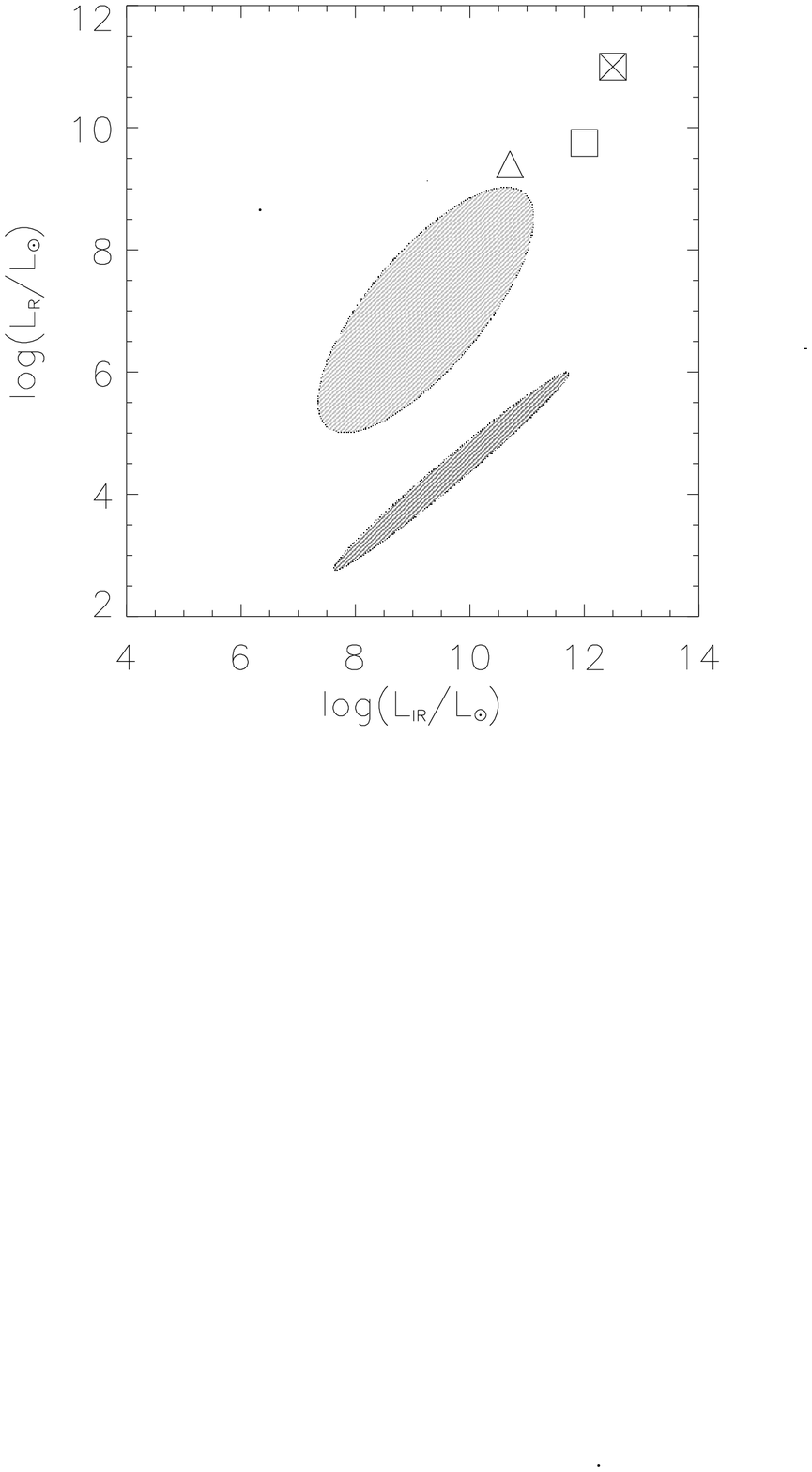}
\caption{Radio versus IR luminosity. Shaded areas represent the regions occupied
by spirals ({\it below}) and low power radio galaxies ({\it above}) from Knapp 
et al. (\cite{knapp}). The triangle
marks the median values for the present sample of radio galaxies, while the 
empty and crossed squares are the low$-z$  and high$-z$ quasars respectively
(Heckman et al. \cite{heckman2})}
\label{knapp}
\end{figure}

We point out  that the median FIR luminosity we find  follows
the extrapolation of the $L_{\rm FIR}$ vs $L_{\rm radio}$ plot for nearby radio 
galaxies presented in Knapp et al. (\cite{knapp}) (see Fig. \ref{knapp}). 

\begin{table*}
\begin{center}
\caption{Luminosities and dust masses for the bright sources }
\begin{tabular}{|l|rr|rr|rr|} \hline
Source& \multicolumn{2}{|c|}{Log(L$_{\rm FIR}$/L$_\odot$)}&
        \multicolumn{2}{|c|}{Hot Component}&
        \multicolumn{2}{|c|}{Cold Component}         \\
      &60--90$\mu$m&174--200$\mu$m&T(K)& Mass($M_\odot)$& T(K)&Mass$(M_\odot)$
 \\       \hline
1345+125     &11.6&12.2 &70 &4.4 $\times 10^6$&   30    &3.5$\times 10^8  $\\
             &    &                &         & & &\\
3C 318       &12.1&--&100 & 8 $\times 10^6$&    --   &-- \\
             &    &                &         & & &\\
1718--64     &9.3&9.5 &60  &2.9$\times 10^4$&    20   &0.1 $\times 10^8$ \\
             &    &                &         & & &\\
1819+39      &12.2& --&65  &1.5$\times 10^7$&    --   & -- \\
             &    &                &         & & &\\
3C 459       &11.9&11.5&70  &4.6 $\times 10^6  $&25    &3.7 $\times 10^8$   \\
             &    &                &         &&&  \\ \hline
\end{tabular}
\label{bright-s}
\end{center}
\end{table*}

\subsection{Temperature and Dust masses}
\label{dust}

On the hypothesis that the dust is transparent at these wavelengths, the MFIR 
radiation due to dust at a uniform temperature $T$ follows a 
modified Planck law given by

\begin{equation}
S(\nu_{0})=\frac{\mu(\nu_{e})B_{\rm b.b.}(T,\nu_{e})M_{\rm dust}(1+z)}{D_{\rm L}^{2}}
\label{eq6.12}
\end{equation}

\noindent
where $\nu_0$ and $\nu_e$ are the observed and emitted frequencies, $\mu$ is the
absorption coefficient, $B_{\rm b.b.}$ is the brightness of a black body at
temperature $T$, $M_{\rm dust}$ is the dust mass, $D_{\rm L}$ is the luminosity distance
and $z$ is the source redshift. The absorption coefficient is rather uncertain.
In order to keep consistency with the values generally assumed in the
literature (e.g.: Hildebrand \cite{hildebrand}; Knapp et al. \cite{knapp}), we take
$\mu(\lambda) = 10 \times (250/\lambda(\mu m))$ for  $\lambda \leq 250 \mu$m.
With respect to other assumed values, this tends to overestimate the
dust mass (e.g. by a factor of two as compared to the $\mu(\lambda)$ assumed
by Hughes et al., \cite{hughes}).

\begin{figure}
\vskip 7 truecm
\includegraphics{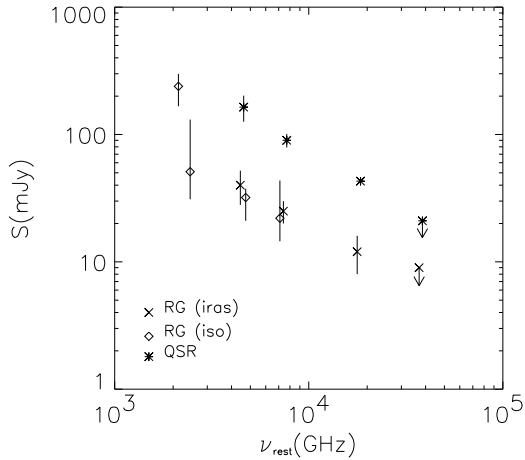}
\caption{IR spectrum of radio galaxies ({\it crosses}: IRAS, {\it diamonds}:
ISO) and quasars from Heckman et al. (\cite{heckman2}) ({\it asterisks}). 
Error bars are 1 $\sigma$ }
\label{spectr}
\end{figure}

The MFIR average spectrum for our sample is  shown in Fig. \ref{spectr}.
In order to describe it better, we used also the following additional upper 
limits or median flux densities: $<$ 5 mJy at 1.3 mm (230 GHz), as 
maximum flux density excess over the extrapolated synchrotron spectrum (Murgia 
et al. \cite{murgia}); $<$ 9 mJy at 12$\mu$m and $(12 \pm 4)$ mJy at 25 $\mu$m 
(Heckman et al. \cite{heckman2}, near sub--sample).

The dependence on frequency of the FIR flux density in the observed wavelength
range 12--200 $\mu$m ($\nu_{\rm rest} \approx 36~000-2000$ GHz), can be described 
by a power law: $$S(\nu) \propto  \nu ^{-1\pm 0.2}$$
At lower frequencies, however, the spectrum of the FIR component must turn over to 
meet 
the 230 GHz upper limit.

This spectrum is clearly not consistent with
a single temperature. A multi--temperature fit (as in models by Sanders et al.
\cite{sanders2}) would give a plausible range of temperatures between $\approx 25 $ K and
$\geq$ 120 K in the wavelength range we have considered here.
However, as the data quality is not sufficient to allow a 
meaningful multi--temperature fit, we have taken a simple two temperature 
approximation, with $T_1 = 80$ K and $T_2 = 25$ K. With corresponding dust 
masses $M_1 \approx 5 \times 10^5~
M_{\odot}$ and $M_2 \approx 2 \times 10^8 M_{\odot}$, we can reproduce
the FIR flux densities  in the range from 60 $\mu$m to 200 $\mu$m.

Any dust component colder than 25 K is poorly constrained by our data. For 
instance, up to $10^9 ~ M_{\odot}$ would be consistent with the 174/200
$\mu$m median flux densities and the 1.3 mm upper limit if the temperature were
$\leq 20$ K. In order to probe such low temperatures, measures at
longer wavelengths would be required, as with SCUBA (Holland et al.,
\cite{holl}). To our knowledge, such data do not exist for our sources.

A comparison of our average mass estimates with those for nearby radio quiet
ellipticals (Bregman et al. \cite{bregman}) and nearby radio galaxies 
(Knapp et al. \cite{knapp})
is not straightforward, as these estimates depend on the wavelengths available 
and on the assumed temperatures.
Bregman et al. (\cite{bregman}) evaluate the temperatures from the 60 $\mu$m to 
100$\mu$m flux density
ratio ($\approx 30$ K) and obtain  $10^5\leq M_{\rm dust}/M_\odot
\leq 10^7$. 
Knapp et al. (\cite{knapp}) assume $T \approx 19$ K and derive  
$10^6\leq M_{\rm dust}/M_\odot \leq 10^8$. 

In our opinion, it is far from being excluded that the range of dust masses in 
these samples and in ours are similar, and the different estimates may
be due to either the different assumed temperatures and/or to the different 
wavelength ranges explored.

\medskip
Using the two temperature approximation, we have also estimated dust masses 
for the brighter sources detected, or possibly detected, at more than one 
wavelength.

The results of this calculations are presented in Table \ref{bright-s}.
Compared to the median values of the total sample,
the sources 1345+145, 3C318, 1819+39
and 3C459 seem to have a higher mass of the warm dust component, 
while the cold dust masses seem to be similar.
As the first three objects are a Seyfert type galaxy, a quasar and a
Broad Line radio galaxy, we may expect that a favorable orientation
of the objects with respect to the line of sight allows us to see
the warmer dust associated with the accretion disk.

\subsection{What heats the dust?}

According to the unification models, the radio galaxies in our sample, which
are very powerful in the radio band and have large luminosities in the 
Narrow Emission Lines ($L_{\rm OIII}\approx 10^{43}$erg s$^{-1}$, Gelderman \& Whittle, 
\cite{gelderman}),  are expected to harbor an obscured quasar (see, e.g., Falcke et al.
\cite{falcke}, in the context of "jet/disk--symbiosis"). A large fraction
of the MFIR emission would then originate from the dust in the absorbing
torus itself, heated by the UV/X radiation emitted at its center, with some
contribution from the interstellar medium, still heated by the central AGN but
now distributed over a scale of several kpc, and therefore with temperatures
decreasing outwards. We  expect this interstellar dust to emit its 
radiation at the longer wavelengths.

If an obscured AGN is present, adopting a heating model as in Sanders et al.
(\cite{sanders2}), we expect dust temperatures in a range 
similar to that estimated in Sect. \ref{dust}. 

If, instead, no AGN is present and heating is due to the stellar radiation
field only, the dust temperature could be $\ltsim$ 30 K, as in nearby radio
quiet elliptical galaxies (Bregman et al. \cite{bregman}). The emission would then
peak at $\lambda \geq 200 \mu$m. An additional contribution to heating could 
arise from a burst of star formation, as may be the case in 3C459.

The interpretation based on heating by an obscured quasar is supported by
the correlation between radio emission and MFIR emission (Knapp et al. 
\cite{knapp};
Heckman et al. \cite{heckman3}), with which  our data are also consistent (Fig.
 \ref{knapp}). Also the strong
luminosity in Narrow Lines, which is often interpreted as powered by 
photo--ionization,
gives support to this interpretation. In addition we note that 
the MFIR average spectrum (for $\lambda_{\rm rest} \geq 20 \mu$m) in nearby radio 
quiet ellipticals (Bregman et al. \cite{bregman}) is 
definitely steeper than that of our radio galaxies and of those in Knapp
et al. (\cite{knapp}). The spectral index of the former is $\alpha_{\rm RQE}
\simeq 2.3 \pm 0.2$, as compared to $\alpha_{\rm us} \simeq 1.0 \pm 0.2$ in our 
sample (Sect. \ref{fir-l}) and $\alpha_{\rm Knapp} \simeq 1.2 \pm 0.2$ in Knapp et al. 
(\cite{knapp}).
This suggests that the heating processes are
different, with the one in radio galaxies being more effective. 

We conclude that our observations are consistent with the MFIR luminosity
in these radio galaxies being powered primarily by the AGN, and assume
that most of the radiation comes from the ``obscuring torus''.

\subsection{Implications for the circumnuclear torus}
\label{impli}

Our average FIR luminosities, consistent with the results of  Heckman et
al. (\cite{heckman2}), are about a factor 4 -- 5 lower than the FIR luminosities of radio 
quasars of comparable radio luminosity (see Fig. \ref{spectr}).

This causes problems for the unification models (see Heckman et al. \cite{heckman2}),
since, if the torus dust is transparent at FIR wavelengths,
its emission should be orientation independent and
the quasars should be as luminous in the FIR as the radio galaxies. On the other 
hand, if the disk is optically thick 
in some wavelength range, one should justify the constant 
flux density ratio between radio galaxies and quasars over a broad wavelength
range (from 12 to 100 $\mu$m).  

Hes et al. (\cite{hes}) discuss  the possibility  that a beamed non--thermal 
component is present in quasars and contributes at the FIR wavelengths.

\subsection{Back to the {\it frustration scenario} for CSS/GPS sources}
The original aim of this work was to study the cold phase of the interstellar
medium  in radio galaxies, in order to see whether its density is high enough
to support the ``frustration scenario" for CSS/GPS sources. 
In this scenario, CSS/GPS and
large size  radio galaxies  have similar lifetimes, so that  the 
difference in size (more than a factor of 100, on average, in our two samples,
Tables \ref{sample} and \ref{sampler}) requires an environment significantly 
denser for the former (De Young  \cite{deyoung}; Fanti et al. \cite{fanti2}).

 The hot component of the ISM is too tenuous and has insufficient pressure to 
 confine the radio source (e.g. O'Dea et al. \cite{odea2};
Readhead et al. \cite{readhead}). Thus the radio source must be confined via
interaction  with relatively cold dense material
along the path of propagation. However, as anticipated in the introduction,
even if a lot of dense cold gas  is present in the host galaxy, distributed in the
disk/torus of the ``unified models'' (which is perpendicular to the radio
axis) it will have no effect on the radio source propagation. 
``Frustration'', if present, must be due to the diffuse cold phase of 
the interstellar gas. 

>From ram pressure computations, the estimates of the density of the
interstellar gas required to frustrate CSS radio galaxies imply masses 
$\geq 2 \times 10^{10} ~ M_{\odot}$ within a volume of $\approx 10$ kpc in 
radius (Fanti et al. \cite{fanti2}), corresponding, for a gas to dust ratio as 
in our own Galaxy, to an excess dust mass $\geq 2 \times 10^8 M_{\odot}$. This
would produce an excess in the FIR emission of the CSS/GPS, as compared to the 
large size radio galaxies, the exact amount depending on the dust temperature. 
The upper
limits to any IR excess derived in Sect. \ref{ave-det} convert into an excess
dust mass of $\ltsim 0.6~10^8 M_{\odot}$ for $T\gtsim 30$ K, (if heating is
due to a powerful AGN), which is not enough for the frustration model.
Even if we attribute the whole MFIR emission we detect from CSS/GPS
to a frustrating medium only, and {\it not} to the dusty disk/torus, the implied
mass (for $T\gtsim$ 30 K), would be still not enough, the upper limits to the 
differences in the MFIR emission being similar to the actual flux densities of
CSS/GPS sources.

As a consequence, we may be confident that there is no evidence
for a  component of the interstellar medium, with temperatures
 $\gtsim$30 K, significantly denser or more massive in CSS/GPS galaxies
 to cause frustration.

For lower temperatures, our FIR observations are not sensitive enough to
firmly constrain larger masses, since the emission would peak at $\lambda > 200~\mu$m,
and the uncertainties we have on the flux density limits at long wavelengths 
are too
large to provide safe constraints. For instance, if the temperature were
$\approx$20 K, up to 5$\times 10^8 M_{\odot}$ of dust would be permitted. This
should be probed by observations at longer IR wavelengths (e.g. with SCUBA).
An additional strong constraint on such large dust masses
is obtained from the color properties of the
host galaxies.  Dust masses as required by frustration, spread over
$\sim$ 10 kpc, would produce absorption in the optical band. 
The implied hydrogen column density, $N_{\rm H} \geq 10^{22}$ g cm$^{-2}$, 
corresponds to an optical depth in the visual $\tau_{\rm v} \approx N_{\rm 
H}/(2~10^{21})$.
Applying a model of homogeneously mixed dust and stars (as in Goudfrooij \& 
de Jong \cite{goudfrooij2}), the above dust mass implies 
$A_{\rm v} \approx 2$ mag., and color reddening E(B--V) $\approx 0.7$, which  is
inconsistent with the data of de Vries et al. (\cite{devries}).

\medskip
If ``frustration" occurs only for GPS (sizes $\leq 1$ kpc), the implied masses
are lower, but their temperature would be larger ($\geq 50$ K, for a 
model as in Sanders et al. \cite{sanders2}), due to the closer proximity to
the hidden quasar, so that the expected IR luminosity would be even more 
discrepant  with what we observe. 

The present conclusion is therefore that the FIR radiation we detect with 
ISO does not imply an amount of gas (heated by the AGN) large enough to 
support the frustration scenario of CSS/GPS radio sources. This is  consistent 
with the results of Owsianik et al. (\cite{owsianik}) and Owsianik \& Conway 
(\cite{owsianik2}), who found fast proper motions of the outer lobes 
in some small symmetric GPS, and with the radio spectra analysis of CSS/GPS by Murgia et al. 
(\cite{murgia}).

\medskip
We finally note that recent observations with SCUBA (Archibald et al. 
\cite{archi}) of high redshift radio galaxies, both CSS and large size ones, 
do not show any significant difference between the two classes of sources.

\bigskip
The MFIR properties of CSS/GPS, however, point out a new problem for the 
alternative interpretation. An important ingredient of the ``youth scenario"  
is that the radio luminosity is required  to decrease with increasing
source size, in order
to explain the numerical proportion of CSS/GPS to large size sources. 
In the current models, this luminosity evolution  is essentially due to 
expansion of the lobe volumes in an external medium of decreasing density 
and not to a
decline of the power carried by the jet with time. CSS/GPS sources would 
evolve, with increasing size, toward lower luminosity larger size radio 
sources. This evolution in luminosity may also be described in terms of a 
higher efficiency for converting the beam power into radio band radiation
(see also Gopal Krishna \& Wiita \cite{gopal}) in the early 
phases of the source evolution.

According to this idea, our two samples (CSS/GPS and comparison), which have
very different sizes but similar radio luminosities,
would have  ``radio engines" of power different by a factor 10 or so (CSS/GPS
being less powerful). As in the past there have been strong suggestions that
the jet power is proportional to the bolometric luminosity of the 
AGN (e.g. Baum \& Heckman \cite{baum2}; Rawlings \& Saunders
\cite{rawlings}; Falcke et al.
\cite{falcke}), one would expect the CSS/GPS to be definitely less powerful
in the MFIR (and in the NL emission). This is not seen in our
observations, at least at 60 and 90 $\mu$m.

Although we have no obvious solution to this problem, we speculate here about
a couple of possibilities.
First there may be an additional source of heating arising from conversion 
of a fraction of the jet power where the jet interacts with the interstellar 
medium. This may be more effective at distances close to the nucleus and 
less so further out as the source ages. The existence of emission line gas
aligned with the radio source in low redshift CSS sources is consistent with
strong interaction between GPS/CSS sources and the dense clouds in the
ISM (de Vries et al. \cite{devries2}). 
Such an effect should roughly compensate for the lower
radiation power from the central engine in CSS/GPS. 

The second possibility is related to the scenario where the radio emission
is triggered by a merger event. We speculate that in the early phase
of the radio source life a large fraction of gas/dust is not fully settled
in a disk/torus but is still distributed over a much larger solid angle, as 
seen from the central continuum source. In this scenario the fraction of
intercepted and reprocessed UV radiation could be much higher in the 
younger/smaller sources, increasing their MFIR emission. We might also
expect the additional obscuration in the smaller sources to result in  
lower emission line luminosities and/or redder colors. This is not
generally seen in GPS and CSS sources, though there are a few cases
which are consistent with this picture (O'Dea \cite{odea3}; de Vries et al.
\cite{devries}). 
In addition, this hypothesis would require the gas/dust to settle into a disk
on a timescale much less than the age of a radio source ($ 10^{7-8}$ yr) 
in order to remove the extra obscuration before the sources propagate to scales
larger than tens of kpc.

At present, both of these explanations appear rather ``ad hoc''. Thus, this
issue remains a problem  for the current evolution models. 

\section{Conclusions} 
\label{concl}
\hskip 6truemm
a) We have presented ISOPHOT observations at $\lambda=$ 60, 90, 174 and 200 $\mu$m
of CSS/GPS radio galaxies and of a matched comparison sample of extended 
radio galaxies. A minority of objects are detected individually at
one or more wavelengths in both samples. 

b) We have co--added the data
for each sample to obtain mean and median flux densities. 
The extrapolated radio spectrum under--predicts the observed MFIR
flux densities, arguing that the MFIR is due to dust rather than to 
synchrotron emission.

c) We find no significant differences in the MFIR flux densities of
the two samples. Our results are then consistent with the CSS/GPS and the
extended  radio galaxies having similar MFIR luminosities. For dust
temperatures $\gtsim$30 K, the deduced masses of the interstellar gas are
lower than required by the frustration scenario. For lower temperatures larger
masses would be allowed for by our data, but they would produce a large amount
of obscuration and reddening in the optical, which is not seen in the existing
data.

All this argues against the CSS/GPS sources being ``frustrated" by a dense 
ambient medium. 

d) Since no significant difference is seen between the two samples, we have combined them
for further analysis. The average $L_{\rm FIR}$ is in the range
$(0.6-1.0)\times 10^{11}$ L$_{\odot}$ or $(2-5)\times 10^{44}$ erg s$^{-1}$.
Over the wavelength range of our observations, the spectrum can be
fitted by a single power law with $\alpha \simeq 1.0 \pm 0.2$.

e) We have fitted simple two temperature ($T_1 = 80$ K and $T_2 = 25$ K)  
models to the IR spectrum  and have derived dust masses of 
$M_1 \approx 5 \times 10^5~ M_{\odot}$ and $M_2 \approx 2 
\times 10^8 M_{\odot}$. The mass of the cold dust appears higher than found
in radio quiet elliptical galaxies, although this difference may 
be due to the differences in the assumed temperatures and the 
sampled rest-frame wavelength coverage. 

f) Our observations are consistent with the MFIR luminosity in these
powerful radio galaxies being 
mostly powered by an obscured AGN, although in some objects a contribution from
star formation is possible.

\begin{acknowledgements}
We are grateful to Dr. M. Haas of MPIA (Heidelberg) for his numerous 
suggestions during the data analysis stage and for providing us with 
specialized software to perform some non standard data reduction. 
CF and FP wish also to thank MPIA staff members for the help obtained during
a short visit at the ISOPHOT group in Heidelberg.

This work was partly supported by the Italian Ministry for University and
Research (MURST) under grant cofin99-02-32
\end{acknowledgements}

\listofobjects
\end{document}